\documentclass[twocolumn,showpacs,aps,prl,superscriptaddress]{revtex4}

\usepackage{graphicx}
\usepackage{dcolumn}
\usepackage{amsmath}

\input pubboard/babarsym

\def\psitwos {\ensuremath{{\psi(2S)}}}

\def\chictwo {\ensuremath{{\chi_{c2}}}}

\def\jpsiee {\ensuremath{{\jpsi\to\epem}}}
\def\jpsimm {\ensuremath{{\jpsi\to\mumu}}}

\def\pstar {\ensuremath{{p^*}}}
\def\cstar {\ensuremath{{\cos\theta^*}}}
\def\hel {\ensuremath{{\theta_H}}}
\def\chel {\ensuremath{{\cos\theta_H}}}
\def\upsresult {\ensuremath{{4.7 \times 10^{-4}}}}
\def\sigmaresult {\ensuremath{{2.52 \pm 0.21 \pm 0.21}}}
\def\hipresult {\ensuremath{{1.87 \pm 0.10 \pm 0.15}}}

\newcommand{\BABARPubYear}    {01}
\newcommand{\BABARPubNumber}  {12}

\newcommand{\SLACPubNumber} {8854}

\def\figurebox#1#2#3{%
    \def\arg{#3}%
    \ifx\arg\empty
    {\hfill\vbox{\hsize#2\hrule\hbox to #2{\vrule\hfill\vbox to #1{\hsize#2\vfill}\vrule}\hrule}\hfill}%
    \else
    {\hfill\epsfbox{#3}\hfill}%
    \fi}

\begin{document}

\preprint{\babar-PUB-\BABARPubYear/\BABARPubNumber} 
\preprint{SLAC-PUB-\SLACPubNumber} 
\preprint{hep-ex/0106044}

\begin{flushleft}
\babar-PUB-\BABARPubYear/\BABARPubNumber\\
SLAC-PUB-\SLACPubNumber\\
hep-ex/0106044
\end{flushleft}

\title{
 Measurement of \jpsi\ production in continuum \epem\ annihilations near
$\sqrt s = 10.6$\gev
}

%
\author{B.~Aubert}
\author{D.~Boutigny}
\author{J.-M.~Gaillard}
\author{A.~Hicheur}
\author{Y.~Karyotakis}
\author{J.P.~Lees}
\author{P.~Robbe}
\author{V.~Tisserand}
\affiliation{Laboratoire de Physique des Particules, F-74941 Annecy-le-Vieux, France }
\author{A.~Palano}
\affiliation{Universit\`a di Bari, Dipartimento di Fisica and INFN, I-70126 Bari, Italy }
\author{G.P.~Chen}
\author{J.C.~Chen}
\author{N.D.~Qi}
\author{G.~Rong}
\author{P.~Wang}
\author{Y.S.~Zhu}
\affiliation{Institute of High Energy Physics, Beijing 100039, China }
% ...
\collaboration{The \babar\ Collaboration}
\noaffiliation

\date{\today}

\begin{abstract}
The production of \jpsi\ mesons in continuum \epem\ annihilations has been
studied with the \babar\ detector at energies near the 
\FourS\ resonance.  The mesons are distinguished from 
\jpsi\ production in $B$ decays
through their center-of-mass momentum and energy. 
We measure the cross section $\epem\to\jpsi X$ to be
\sigmaresult\pb.
We set a 90\% CL upper limit on the branching fraction
for direct $\FourS\to\jpsi X$ decays at \upsresult.
\end{abstract}


\maketitle

The development of non-relativistic QCD (NRQCD) represents a
significant advance in the theory of the production of heavy
quarkonium (\qqbar) states \cite{ref:bodwin}.  
In particular, it provides an explanation \cite{ref:psi2sbrat}
for the cross section for \psitwos\ production observed by CDF
\cite{ref:abe}, which is a factor of 30 larger than expected
from previous models.  The
enhancement is attributed to the production of a \ccbar\ pair in a
color octet state, which then evolves into the charmonium (\ccbar)
meson along with other light hadrons.  A similar contribution is expected
in NRQCD 
for \jpsi\ production in \epem\ 
annihilation \cite{ref:yuan, ref:schuler}, but
is absent in the color singlet model \cite{ref:cho}.

Significant continuum \jpsi\ production---as distinct from production
in $B$ decay at the \FourS\ resonance---has not been observed
previously in \epem\ annihilation below the $Z$ resonance. It
therefore represents a good test of NRQCD.  
In particular, matrix elements extracted from different \jpsi\
production processes should be consistent
\cite{ref:leibovich}.  In addition, momentum,
polarization and particularly the angular distributions of the \jpsi\
distinguish between theoretical approaches \cite{ref:braaten}.
Despite NRQCD's successes, it is not clear
that it correctly explains \cite{ref:polbraaten} the CDF measurements
of \jpsi\ polarization \cite{ref:affolder}, or measurements of
\jpsi\ photoproduction at HERA \cite{ref:hera, ref:cacciari}.

The study reported here uses
20.7\invfb\ of data collected at the \FourS\
resonance (10.58\gev) and 2.59\invfb\ collected at 10.54\gev, below
the threshold for \BB\ creation.  The luminosity-weighted
center-of-mass (CM) energy is 10.57\gev.  

The data were collected with the \babar\ detector \cite{ref:babarnim} 
located at the
PEP-II collider at the Stanford Linear Accelerator Center. PEP-II
collides 9\gev\ electrons with 3.1\gev\ positrons to create
a center of mass moving along the $z$
axis with a Lorentz boost of $\beta\gamma = 0.56$.

The momenta and trajectories of charged particles are reconstructed
with two detector systems located in a 1.5-T solenoidal magnetic
field: a five-layer, double-sided silicon vertex tracker (SVT) and a
40-layer drift chamber (DCH).  The fiducial volume covers the polar
angular region $0.41 < \theta < 2.54$~rad, 86\% of the solid angle in the
CM frame. 

The energies of electrons and photons are measured in a
CsI(Tl) electromagnetic calorimeter (EMC) in the
fiducial volume $0.41 < \theta < 2.41$~rad, 84\% of the solid angle in the
CM frame.  
The instrumented flux return (IFR) is used to detect muons.
The DIRC, a unique
Cherenkov radiation detection device, distinguishes charged particles
of different masses.

\jpsi\ mesons are reconstructed via decays to electron or muon 
pairs.  The leptons
must form high-quality tracks with 
$0.41 < \theta < 2.41$~rad: they must have $p_t > 0.1$\gevc\ and
momentum below 10\gevc, have at least 12 hits in the DCH, and
approach within 10\cm\ of the beam spot in $z$ and within 1.5\cm\
of the beam line.  The beam spot rms size is approximately 0.9\cm\ in
$z$, 120\mum\ horizontally and 5.6\mum\ vertically.

One electron candidate must have an energy deposit in the EMC of at
least 75\% of its momentum.  The other must have between 89\% and
120\%, and must also have an energy deposition in the DCH and a signal 
in the DIRC consistent with expectations for an electron.
Both
must satisfy criteria on the shape of the EMC deposit.  If possible,
photons radiated by electrons traversing material prior to the DCH are
combined with the track.

Muon candidates must deposit less than 0.5\gev\ in the EMC
(2.3 times the minimum-ionizing peak), 
penetrate at least
two interaction lengths $\lambda$ of material, and have a pattern of
hits consistent with the trajectory of a muon.
We require that the material traversed by one candidate be within
1~$\lambda$ of that expected for a muon; for the other candidate, this 
is relaxed to 2~$\lambda$.

The mass of the \jpsi\ candidate is calculated after constraining
the two lepton candidates to a common origin.

To reject interactions with residual gas in the
beam pipe or with the beam pipe wall, we 
construct an event vertex using all tracks in the
fiducial volume and require it to be located within 6\cm\ of the beam
spot in $z$ and within 0.5\cm\ of the beam line.  To suppress a
substantial background from radiative Bhabha ($\epem\gamma$) events in
which the photon converts to an \epem\ pair, five tracks are required
in events with a \jpsiee\ candidate.

At this point, the data includes \jpsi\ mesons both from 
our signal---continuum-produced
\jpsi\ mesons and \jpsi\ mesons from the decay of continuum-produced
\psitwos\ and $\chi_{cJ}$ mesons---and from other known sources.
We apply additional selection criteria to suppress these other sources
based on their kinematic properties.

The most copious background, $B\to \jpsi X$, is eliminated by
requiring the
\jpsi\ momentum in the CM frame (\pstar) to be greater
than 2\gevc, above the kinematic limit for $B$ decays.  This
requirement is dropped for data recorded below the \FourS\ resonance.

Other background sources include initial-state radiation (ISR)
production of \jpsi\ mesons, $\epem \to \gamma\jpsi$, or of the
\psitwos, with $\psitwos \to \jpsi X$.  ISR production of lower-mass
$\Upsilon$ resonances is negligible.  Two photon production of the
\chictwo\ can produce \jpsi\ mesons via $\chictwo \to \gamma \jpsi$.
Because the out-going electron and positron are rarely reconstructed,
this process, like the ISR \jpsi\ production, contains only two
tracks.  We therefore require three high-quality
tracks with $0.41<\theta < 2.54$~rad.

\begin{figure}
\includegraphics[width=\linewidth]{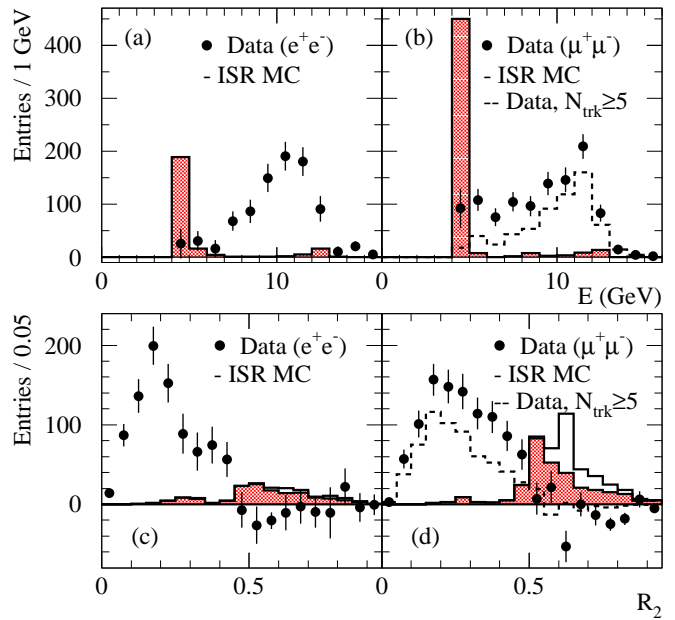}
\caption{\jpsi\ signal events observed as a function
of visible energy $E$ 
in the 
(a) \epem\ and (b) \mumu\ final states; 
$R_2$ distribution for (c) \epem\ and (d) \mumu.
The histogram is the predicted ISR background that has been
subtracted from data; the filled histogram is the ISR \psitwos\
component only. 
A requirement of $\ge 5$ tracks is applied to the
\epem\ sample; applying it to the \mumu\ sample produces the 
dashed histogram.
Event preselection requires $E>4$\gev\ and $R_2<0.95$.}
\label{fig:etotr2}
\end{figure}

The remaining background is primarily ISR \psitwos\ decays to
\jpsi\pipi, plus some ISR \jpsi\ events in which the ISR photon
converts.  To suppress these, we require the visible energy $E$ to be
greater than 5\gev, and the ratio of the second to the zeroth
Fox-Wolfram moment \cite{ref:fox}, $R_2$, to be less than 0.5.  Both
are calculated from tracks and neutral clusters in the fiducial
volume.  Figure~\ref{fig:etotr2}, which displays the visible energy
and $R_2$ distributions for our signal and for simulated ISR
background, motivates these criteria.

The ISR distributions in Fig.~\ref{fig:etotr2} are obtained from a
full detector simulation.  All selection criteria are applied, other
than the one on the quantity being plotted.
ISR kinematics ensures $E<5$\gev\ when the
photon is outside the fiducial volume unless it 
interacts in material and deposits additional energy in the detector.
The rate
of such interactions is not accurately simulated and so is obtained by
a comparison to data for $E<5$\gev.  
Approximately 3.5\% of the \jpsi\ meson events that satisfy all
criteria are from this background; 
an additional
$\sim 1.6$\% are ISR events with the photon in the fiducial volume.
Systematic errors on the remaining backgrounds are estimated from a
comparison between simulation and data for $E< 5$\gev\ and for events
in which the ISR photon is reconstructed.

\jpsi\ production as a function of $E$ is
obtained in data by fitting the dilepton mass distribution in 1-GeV wide
energy intervals after applying all other selection criteria. The fit
uses a polynomial function for the background distribution.
The \jpsi\ mass function is obtained from a complete
simulation of $B\to \jpsi X$ events,  
convolved with a Gaussian
distribution to match the resolution of
12\mevcc\ observed in data in a sample of approximately 14,000 
$B\to\jpsi X$ events. 
The signal distribution in $E$ is obtained by subtracting the ISR backgrounds 
from the data distribution.

A similar process is
used for $R_2$.  
Figures~\ref{fig:etotr2}(c) and (d) show 
there is little signal
above $R_2$ of 0.5.  In this respect, the continuum \jpsi\ events 
are more similar
to \BB\ events, in which the energy is distributed
spherically, than $\ccbar$
events, which tend to be jet-like.

\begin{figure}
\includegraphics[width=\linewidth]{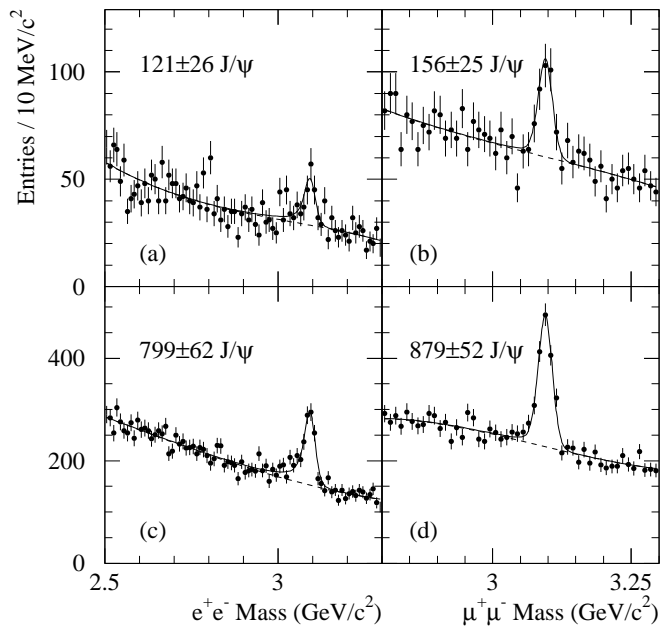}
\caption{Mass distribution of \jpsi\ candidates reconstructed in data
recorded below the \FourS\ resonance in the (a) \epem\ and (b) \mumu\
final states.  Mass distributions for $\pstar > 2$\gevc\ in data at the
\FourS\ resonance in (c) \epem\ and (d) \mumu\ final states. 
The number of \jpsi\
mesons extracted by a fit to the distribution is shown on each graph.}
\label{fig:yield}
\end{figure}

The mass distributions of the selected \jpsi\ candidates show clear
signals for both \epem\ and \mumu\ final states, both on and below
resonance (Fig.~\ref{fig:yield}).

To determine the production cross section, we perform
mass fits in 15 \pstar\ - \cstar\ bins, where $\theta^*$ is the
polar angle of the candidate
in the CM frame.  This allows us to correct
for the variation of efficiency with \pstar\ 
and \cstar.
The cross section is given by:
\begin{equation}
\sigma_{\epem\to\jpsi X} = 
\sum_{i,j} \frac{ \left( N_{ij} - B_{ij} \right ) }{
 \epsilon^R_{ij} \cdot \epsilon^E \cdot 
 \BR_{\jpsi\to\ellell} \cdot \lum_i}, 
\end{equation}
where the sum is over three \pstar\ ($i$) 
and five \cstar\ ($j$) bins. $N_{ij}$
is the number of \jpsi\ mesons in the bin, where electrons and muons
are analyzed separately, but off and on-resonance data are combined.
The sum of the yields from the 15 fits agrees to within 1\% with the 
yields in Fig.~\ref{fig:yield}.
$B_{ij}$ is the ISR background,
$\BR_{\jpsi\to\ellell}$ is the $\jpsi\to\epem$ or \mumu\ 
 branching fraction \cite{ref:pdg2000}, and 
$\lum_i$ is the integrated luminosity---sum of on
plus off-resonance for $\pstar>2$\gevc, off-resonance only for
$\pstar<2$\gevc.

The reconstruction efficiency $\epsilon^R_{ij}$ (acceptance, track
quality and lepton identification) is calculated 
in each bin with simulated
unpolarized \jpsi\ mesons uniformly distributed in \pstar\ and \cstar.
The efficiency decreases with increasing \pstar\ and
\cstar\ due to acceptance. The average $\epsilon^R_{ij}$ is
0.63 for \jpsiee\ and 0.48 for \jpsimm, where the difference is due to 
lepton identification.

Particle identification efficiency is
verified in data by comparing the number of \jpsi\ mesons in $B$ decays 
in which one or both leptons satisfy the
requirements.  The efficiency of the track-quality selection
is studied by comparing tracks found in the SVT and DCH.

The components of $\epsilon^E$, the event selection efficiency, are
determined as follows.  
We estimate the efficiency of the requirements on the number of
high-quality tracks, primary vertex location, and total energy 
to be the average of simulated \ccbar\ and
\BB\ events, and the uncertainty to be one-half the difference.
We use \BB\ events for $R_2$.  

The efficiency of the five track requirement applied to \epem\
candidates is 0.67, 
obtained by comparing the net \jpsi\ yield, in both
\epem\ and \mumu\ final states,  
in events
passing and failing the requirement.
Overall, $\epsilon^E = 0.59$ for \epem\ and
0.89 for \mumu.

The calculations of the $\jpsi X$ cross section from the \epem\ and
\mumu\ final states are consistent: the ratio
$\sigma(\mumu)/\sigma(\epem)$ is $0.93 \pm 0.11$ for $\pstar >2$~\gevc.  
The two values
are combined, accounting for common systematic errors, to obtain
\begin{equation}
\sigma_{\epem\to\jpsi X} = \sigmaresult \text{ pb,}
\label{eqn:result}
\end{equation}
where the first error is statistical and the second systematic.
With existing values for matrix elements, color singlet cross
section estimates range from 0.45 to 0.81\pb\
\cite{ref:yuan, ref:schuler, ref:cho}, while NRQCD cross sections,
including a color octet component, range from 1.1 to 1.6\pb\
\cite{ref:yuan, ref:schuler}.

The dominant component of the 
8.3\% systematic error is a 7.2\% uncertainty on
$\epsilon^E$ common to both the \epem\ and \mumu\ cases and a 4.9\%
uncertainty due to the five track requirement.  Other contributions
include 2.4\% due to track quality cuts; 1.5\% from the luminosity;
1.8\% (electrons) or 1.4\% (muons) from particle identification; and
1.2\% from the ISR background.  

The statistical error is dominated by the uncertainty on the
contribution below
\pstar\ of 2\gevc.  Restricting the measurement to $\pstar>2$\gevc\ gives
$\sigma_{\epem\to\jpsi X} = \hipresult$\pb.

In determining
the cross sections, we assume that there are no
\jpsi\ mesons from direct \FourS\ decays.
We quantify this statement using the $\pstar>2$\gevc\ component. 
We scale the off-resonance event yield to the on-resonance luminosity
and subtract it from the on-resonance yield.
The excess, attributable to \FourS\ decays, is consistent with zero:
$-120 \pm 179$ \epem\ events and $176 \pm 138$ \mumu, in a sample of
$\left( 22.7 \pm 0.4 \right )\times 10^6$ \FourS\ decays.  Using the
average reconstruction efficiency for $\pstar>2$\gevc\ (0.62 for
\epem\ and 0.45 for \mumu), we obtain $\BR_{\FourS\to\jpsi X} =
\left( 1.5 \pm 2.2 \pm 0.1 \right ) \times 10^{-4}$.
A Bayesian 90\% confidence level upper limit with a uniform prior
above zero is:
\begin{equation}
\BR_{\FourS\to\jpsi X} < \upsresult
\text{ (90\% CL),}
\end{equation}
for \jpsi\ with $\pstar> 2$\gevc.
This result disagrees with 
a previous publication \cite{ref:alexander}. 
In NRQCD, the expected partial width is similar to that 
for the $\Upsilon(1S)$ \cite{ref:schuler, ref:cheung}, implying a
branching fraction of a few $\times 10^{-6}$.
Note that a true branching fraction of $10^{-4}$ would correspond to
an effective cross section of 0.10~\pb.

\begin{figure}
\includegraphics[width=\linewidth]{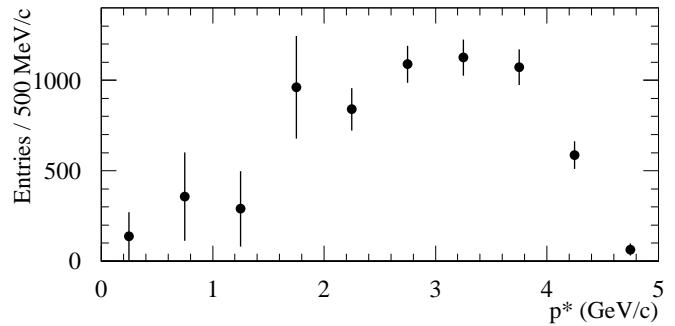}
\caption{Center of mass momentum distribution of \jpsi\ mesons
produced in continuum \epem\ annihilation.}
\label{fig:pstar}
\end{figure}

Production and decay properties of the \jpsi\ have also been studied.
The \pstar\ distribution is obtained by dividing the sample into
500\mevc\ wide intervals, fitting the resulting mass distribution,
subtracting predicted ISR backgrounds, correcting for the
reconstruction efficiency, and normalizing for different luminosities
(Fig.~\ref{fig:pstar}).  

\begin{figure}
\includegraphics[width=\linewidth]{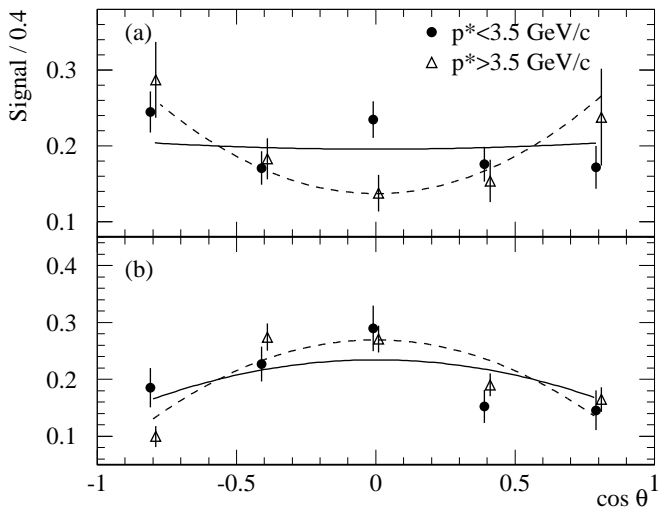}
\caption{(a) Production angle (\cstar) 
distribution for \jpsi\ mesons produced in
continuum \epem\ annihilation; (b) helicity (\chel) distribution.
Solid curve is the fit to $\pstar < 3.5$\gevc; dashed curve is for
$\pstar > 3.5$\gevc.}
\label{fig:cstar}
\end{figure}

The distribution of the signal in \cstar\ has been extracted  
and
fit with $1 + A\cdot \cos^2\theta^*$.
Both NRQCD and color singlet 
calculations predict a flat distribution ($A \approx 0$) at low
\pstar.  At high momentum, NRQCD predicts
$0.6 < A < 1.0$ while the color singlet model predicts $A \approx
-0.8$ \cite{ref:braaten}.  We measure the distribution separately for
low and high momentum mesons, selecting $\pstar = 3.5$\gevc\ as the
boundary.  We proceed as for the \pstar\ distribution, with mass fits
performed in \cstar\ intervals of width 0.4.  The distributions are
then normalized to unit area (Fig.~\ref{fig:cstar}a).  We find $A =
0.05 \pm 0.22$ for $\pstar < 3.5$\gevc\ and $A = 1.5 \pm 0.6$ for
$\pstar > 3.5$\gevc, clearly favoring NRQCD.

Finally, we obtain the helicity angle \hel\ distribution for 
the two \pstar\ ranges by fitting
mass distributions in intervals of width 0.4 in \chel\ 
(Fig.~\ref{fig:cstar}b). 
The helicity is the angle, measured in the
rest frame of the \jpsi, between the positively charged lepton
daughter and the direction of the \jpsi\ measured in the 
CM frame.  Fitting the function 
$3\left( 1 + \alpha\cdot \cos^2 \theta_H
\right ) / 2 \left(\alpha + 3 \right )$, we obtain a \jpsi\
polarization $\alpha = -0.46
\pm 0.21$ for $\pstar<3.5$\gevc\ and $\alpha = -0.80 \pm 0.09$ for
$\pstar>3.5$\gevc.
$\alpha = 0$ indicates an
unpolarized distribution, $\alpha = 1$ transversely polarized, and
$\alpha = -1$ longitudinally polarized.


In summary, we measure the cross section 
$\sigma_{\epem\to\jpsi X} = \sigmaresult$\pb.  Restricting to
$\pstar > 2$\gevc, we find 
\hipresult\pb. The total cross section
and the angular distribution favor the NRQCD calculation
over the color singlet model. 
We set a 90\% CL upper limit on the branching fraction 
$\FourS\to\jpsi X$ of \upsresult.

We are grateful for the excellent luminosity and machine conditions
provided by our \pep2\ colleagues, 
and for the substantial dedicated effort from
the computing organizations that support \babar.
The collaborating institutions wish to thank 
SLAC for its support and kind hospitality. 
This work is supported by
DOE
and NSF (USA),
NSERC (Canada),
IHEP (China),
CEA and
CNRS-IN2P3
(France),
BMBF and DFG
(Germany),
INFN (Italy),
NFR (Norway),
MIST (Russia), and
PPARC (United Kingdom). 
Individuals have received support from the 
A.~P.~Sloan Foundation, 
Research Corporation,
and Alexander von Humboldt Foundation.

\end{document}